# Sequential superconductor-Bose insulator-Fermi insulator phase transitions in two-dimensional *a*-WSi

Xiaofu Zhang and Andreas Schilling

Physik-Institut, Universität Zürich, Winterthurerstrasse 190, Zürich CH-8057, Switzerland.

A zero-temperature magnetic-field-driven superconductor to insulator transition (SIT) in quasi-two-dimensional superconductors is expected to occur when the applied magnetic-field crosses a certain critical value [1,2]. A fundamental question is whether this transition is due to the localization of Cooper pairs or due to the destruction of them. Here we address this question by studying the SIT in amorphous WSi. Transport measurements reveal the localization of Cooper pairs at a quantum critical field $B_c^1$ (Bose-insulator), with a product of the correlation length and dynamical exponents $zv \sim 4/3$ near the quantum critical point (QCP). Beyond $B_c^1$, superconducting fluctuations still persist at finite temperatures. Above a second critical field $B_c^2 > B_c^1$, the Cooper pairs are destroyed and the film becomes a Fermi-insulator. The different phases all merge at a tricritical point at finite temperatures with $zv = 2/3$. Our results suggest a sequential superconductor to Bose insulator to Fermi insulator phase transition, which differs from the conventional scenario involving a single quantum critical point.

Zero-temperature quantum phase transitions (QPTs) between different quantum states are expected to occur when a control parameter crosses over a certain critical value [1-3]. The zero-temperature superconductor-to-insulator QPT in two dimensions, for example, is a well-documented manifestation for such a QPT [2], which can be tuned by control parameters such as external magnetic field, electrical field, charge density, disorder, or thickness [2,4-7]. In the vicinity of the QCP, a quantum critical region is formed due to quantum fluctuations, and the physical behaviors of QPTs can be perceived through measurements at nonzero temperature within the quantum critical region [2].

The superconducting state is characterized by an order parameter, in terms of amplitude (related to energy gap $\Delta$ or Cooper-pair density $n_s$), and phase $\phi$. As a result, a SIT can be grouped into two distinct classes, with fermionic and with bosonic descriptions. The fermionic scenario describes the SIT as a result of the amplitude fluctuations, in which a SIT is driven by the breaking of Cooper pairs and the localization of electrons, forming a Fermi insulator [8-11]. On the contrary, the SIT in the bosonic description is driven by fluctuations of the phase, in which the superconducting and the insulating states with different symmetry are separated by a QCP [12-14]. On the superconducting side of such a transition, the Cooper pairs are mobile. Magnetic vortices are localized and can be bound into vortex-antivortex pairs [4]. On the insulating side, the vortices are mobile but the Cooper pairs are localized, forming a Bose-insulating state [13-16]. A number of SIT experiments suggest that the localization of Cooper pairs leads to a quantum-percolation induced superconductor-to-Bose-insulator transition, with a product of the correlation length and dynamical exponents $z\nu \sim 4/3$ near the quantum critical point [2,4,17-21]. However, a superconductor-to-Fermi-insulator transition has also been found in some experiments, with $z\nu \sim 2/3$ [5,6,22,23]. Although many SIT experiments driven by different control parameters have been reported for a variety of materials and with varying $z\nu$ products, the nature of the non-superconducting ground state at zero temperature and the reason for the difference in the respective $z\nu$ values are still unclear [2,24].

Here we demonstrate that a perpendicular magnetic field drives a 2D $W_{0.75}Si_{0.25}$ film (with thickness 5 nm, see our Ref. 25) into an insulating state in two stages, as it is shown in Fig. 1. At a critical field $B_c^1 = 6.25$ T, the sheet resistance $R_s$ is independent of temperature in the zero-temperature limit, which is a hallmark of a QPT, and the superconducting state is driven into an insulating state. Above $B_c^1$, namely on the insulating side of the QPT, pronounced superconducting fluctuations still exist in the system, persisting up to $T \sim 5$ K, well above the zero-field critical temperature $T_c \sim 3.95$ K of the film, as it is shown in Fig. 1(a) and (b). With increasing the magnetic field further, the superconducting fluctuations gradually diminish, and a second classical transition occurs at a critical field $B_c^2 = 7.3$ T, above which the superconducting fluctuations disappear (see Fig. 1(c)). We will interpret the classical critical field $B_c^2$ as a transition to a Fermi-insulator.

An experimental characteristic for such transitions is the scaling behavior of physical quantities [1,2]. When the system enters the critical region close to the phase transition, the measured physical quantities show a power-law dependence on the rescaled spatial and time coordinates, *i.e.*, the correlation length $\xi$ (depending on the proximity to the phase transition point, $\xi \sim |x - x_c|^{-\nu}$) and correlation time $\tau$ ($\tau \propto \xi^z \propto |x - x_c|^{-z\nu}$). These two scaling parameters $\xi$ and $\tau$, which both diverge at the transition point, are determined by the microscopic parameters in the Hamiltonian describing the system. The correlation length exponent $\nu$ and dynamical critical exponent $z$, as well as the scaling functions, however, are determined only by the universality class of phase transition, which depends only on the general properties of the system, such as the space dimensionality *d*, presence of disorder, and the symmetry of the order parameter manifold [2-4]. For a magnetic-field driven SIT, the sheet resistance can be rescaled as $R_s(B,T) \propto |B - B_c|^{2-d} F(|B - B_c|T^{-1/z\nu})$, with an arbitrary but for the system universal scaling function $F(x)$ with $F(0) = 1$. In the 2D (*d* = 2) limit, $\xi$ drops out, $R_s$ at the critical point becomes universal ($R_Q$), and $R_s(B,T) = R_Q F(|B - B_c|T^{-1/z\nu})$, where $R_Q = h/4e^2$ is the quantum resistance for Cooper pairs [3].

Figure 2(a) shows the magnetoresistance $R_s$ data for temperatures ranging from 0.34 to 0.54 K. The $R_s(B)$ curves clearly exhibit a crossing point at a critical magnetic field ($B_c^1 = 6.25$ T, $R_c^1 = 399.7$ Ω). It corresponds to the separatrix with a constant $R_s(T)$ in the zero-temperature limit, thereby defining the QCP in Fig. 1(b). This critical point separates the insulating state from the superconducting state, in that $dR_s/dT$ changes its sign from superconducting behavior to insulating behavior at different sides of $B_c^1$. In the critical regime, the $R_s(B,T)$ data should collapse onto a single curve by scaling the abscissa as $|B - B_c^1|t$, where $t = T^{-1/z\nu}$ [2,24]. Here we set $t$ at the lowest investigated temperature as $t(0.34 \text{ K}) = 1$, and all the other data are then collapsed onto $R_s(B, T = 0.34 \text{ K})$ by adjusting $t$ for each temperature $T$, as it is shown in Fig. 2(b). As a result, the critical exponent $z\nu$ can be retrieved by a power-law fit of $t(T)$, which yields $z\nu =$

1.33~4/3 with high precision on both sides of the SIT (Fig. 2(b), inset). Alternatively, a critical exponent $zv \sim 4/3$ can also be obtained from scaling other parameters, such as $T_c(B)$. By carefully checking the $T_c(B)$ dependence in the zero-temperature limit, the QCP $(B_c^1, T = 0)$ is very close to the extrapolated ending point of the $T_c(B)$ curve (see Fig. 4), which is consistent with theoretical expectations [17]. According to general scaling arguments [3,23], the thermodynamic superconducting transition temperature $T_c(B)$ scales in the vicinity of the QCP according to $T_c(B) \propto |B - B_c^1|^{zv}$, which we indeed observe with $zv \sim 4/3$ (see supplementary material S1).

Above the QCP $B_c^1$, superconducting fluctuations still persist in the system at finite temperatures (Fig. 1(b)), but the system is a zero-temperature insulator. At finite temperatures, the $R_s(T)$ data (Figs. 1(b) and (c)) show a minimum, separating the fluctuations dominated region ($dR_s(T)/dT > 0$) from the insulating region ($dR_s(T)/dT < 0$), thereby defining a phase boundary between the Bose-insulating and the fluctuation dominated normal state (see Fig. 4). Such finite temperature superconducting fluctuations beyond the superconductor to insulator QPTs have also been observed in amorphous InO$_x$ systems [17], liquid helium-quenched gallium films [26], and quench-condensed ultrathin beryllium films [20]. The superconducting fluctuations above $B_c^1$ are gradually weakened by increasing the magnetic field until a second critical field ($B_c^2 = 7.3$ T, $R_c^2 = 415.3$ Ω) is reached, beyond which the superconducting fluctuations disappear (Fig. 1(c)). Inspecting the corresponding $R_s(B,T)$ data in the temperature range from 3 to 5 K (Fig. 3(a)) reveals another crossing point at $B_c^2$, which we interpret as the critical value for Cooper-pair breaking, i.e., the transition to a Fermi insulating state. The scaling analysis in the critical regime of $B_c^2$, analogous to that we applied at $B_c^1$, shows an excellent collapse of the $R_s(B,T)$ curves (Fig. 3(b)), leading to $zv = 0.67 \sim 2/3$ (Fig. 3(b), inset).

It is natural to interpret our observations in the following way. The perfect scaling behavior down to zero temperature at $B_c^1$ is clearly consistent with the bosonic description, where Cooper pairs exist on both sides of the superconductor-to-insulator QPT which is driven by quantum phase fluctuations [2-4,13,14,24]. The zero-temperature insulating state is characterized by the loss of global phase coherence. However, superconducting amplitude fluctuations still exist on the insulating side above $B_c^1$ [27,28]. The physical picture on the insulating side involves localized Cooper pairs and freely moving vortices, forming a bosonic insulating state [2,13,14]. The product $z\nu = 4/3$ is the simplest and most direct manifestation of such a superconductor-to-Bose insulator QPT since it is only depends on the universality class of system. By assuming $z = 1$, $\nu \sim 4/3 > 1$ corresponds to the $T \to 0$ SIT in 2D disordered systems, as observed in some conventional amorphous superconductors [18,29], two-dimensional electron gases at oxide interfaces [24], ultrathin high-$T_C$ superconductors [4,21], or more recently, in graphene-metal hybrids [30]. The critical sheet resistance $R_c \approx 400\ \Omega$ measured in our amorphous WiSi film at $B_c^1$ is one order of magnitude smaller than $R_Q \approx 6\ k\Omega$, however. We note that certain deviations of experimental $R_c$ from the ideal theoretical value $R_Q$ are not uncommon, and can be ascribed to the contribution of a Fermionic channel to the total resistance [18].

Above $B_c^1$ on the insulating side, the sizeable superconducting fluctuations must be attributed to localized Cooper pairs in superconducting islands. In disordered or even amorphous superconductors which can be rendered into an inhomogeneous state by order-parameter amplitude fluctuations, such superconducting islands can appear in the system in high magnetic fields [2,27,28,31]. At a critical field $B_c^2$, the localized Cooper pairs are completely destroyed, and a $z\nu \sim 2/3$ classical transition to Fermi-insulator occurs. Such transitions with $z\nu \sim 2/3$ and driven by a perpendicular magnetic field have also been observed in other conventional 2D films, such as $a$-NbSi [22], $a$-bismuth [32], or in La$_{2-x}$Sr$_x$CuO$_4$ films [21]. A comparison of our $a$-WSi with $a$-NbSi, $a$-bismuth, and La$_{2-x}$Sr$_x$CuO$_4$ ($H_1^*$ in Ref. 21), reveals a very similar behavior with respect to the classical critical field $B_c^2$, in which the critical regions are actually located near $T_c(0)$. Other SITs with $z\nu \sim 2/3$ have been found in electrostatically tuned superconducting LaTiO$_3$/SrTiO$_3$ or LaAlO$_3$/SrTiO$_3$ interfaces [23,24]. At the LaTiO$_3$/SrTiO$_3$ interfaces, the global superconductivity is supposed to be determined by the presence of superconducting islands coupled by non-superconducting metallic regions, and the magnetic field driven SIT with $z\nu \sim 2/3$ corresponds to the vanishing of Cooper pairs within the islands [24]. At the electrical field stimulated LaAlO$_3$/SrTiO$_3$ superconducting interfaces, the charge density driven superconductor to Fermi insulator transition with $z\nu \sim 2/3$ at large negative electrical field can be ascribed to the depletion of carriers and breaking of Cooper pairs [23].

The scenarios in which SITs with $z\nu \sim 2/3$ correspond to a phase transition associated with pair breaking are further supported by the magnetic field tuned SITs with varying bias current in 2D beryllium films [20]. At negligible bias current, an SIT with $z\nu \sim 4/3$ is observed, while with a dc current near the zero-field depairing critical currents, the $z\nu$ is driven from $\sim 4/3$ to $\sim 2/3$ [20].

The phase transitions among these distinct states in our WSi films are visualized in the schematic phase diagram in Fig. 4. At low field, the superconducting state and the fluctuation-dominated normal region are separated by the $T_c(B)$ line, which we experimentally determined by a $R_n/2$ standard. This $T_c(B)$ line terminates at the QCP $B_c^1$ in the zero-temperature limit as $T_c(B) \propto |B - B_c^1|^{z\nu}$. At zero temperature, the superconducting and the Bose insulating quantum ground state are separated by the QPT at $B_c^1$. Above $B_c^1$ and at elevated temperatures, finite temperature classical transitions between the fluctuation-dominated normal and the Bose insulating states are observed. They end at the second critical field $B_c^2$ and at a tricritical point around $T_{\text{tcp}} \approx T_c(0) \approx 4$ K. At temperatures above $T_{\text{tcp}}$, the fluctuation-dominated state directly crosses over into the Fermi insulating state. At low temperatures, the localized Cooper pairs within the superconducting islands in the Bose insulating state are subsequently destroyed by the increasing field, and a Bose insulator to Fermi insulator transition occurs. By carefully comparing the response of the $R_s(B)$ data to the magnetic field near $B_c^1$ and $B_c^2$, we state that the crossover between Bose insulating (strong field dependence) and Fermi insulating state (weak field dependence) must correspond to a virtually horizontal line terminating at the tricritical point at $T_{\text{tcp}}$ (Fig. 4, supplementary material S2), which characterizes the field $B_c^2$ as the depairing field of the Cooper pairs.

**Acknowledgments**: We acknowledge A. Goldman, T. Neupert, J. Chang, Q. Wang and D. Destraz for discussions. This work was supported by UZH and Swiss National Foundation Project Nr. 20-175554.

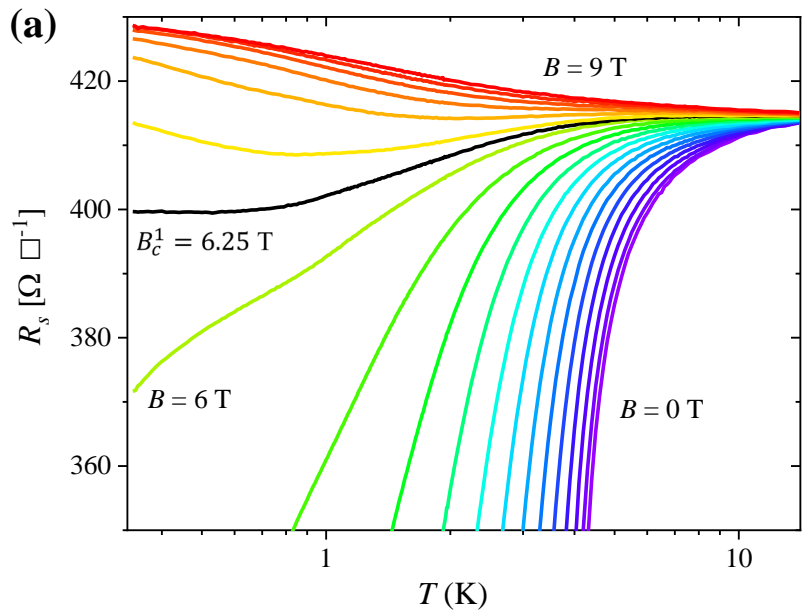

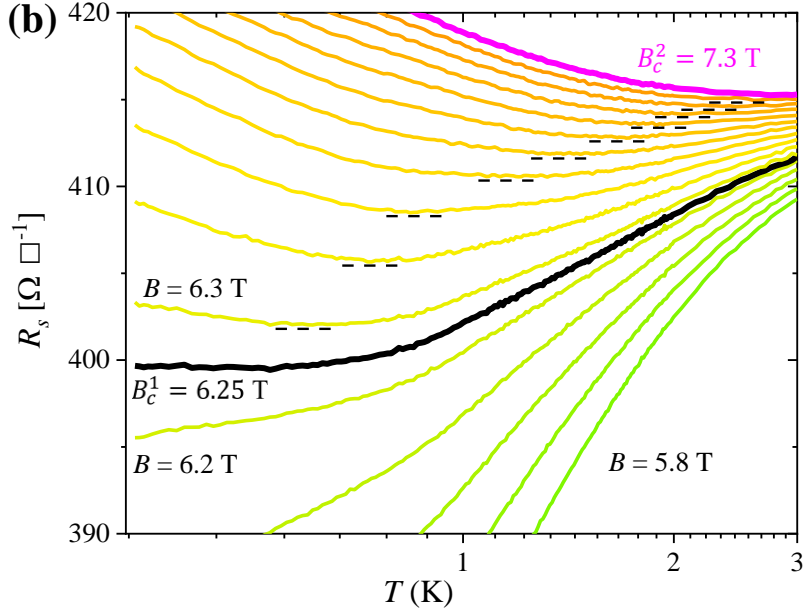

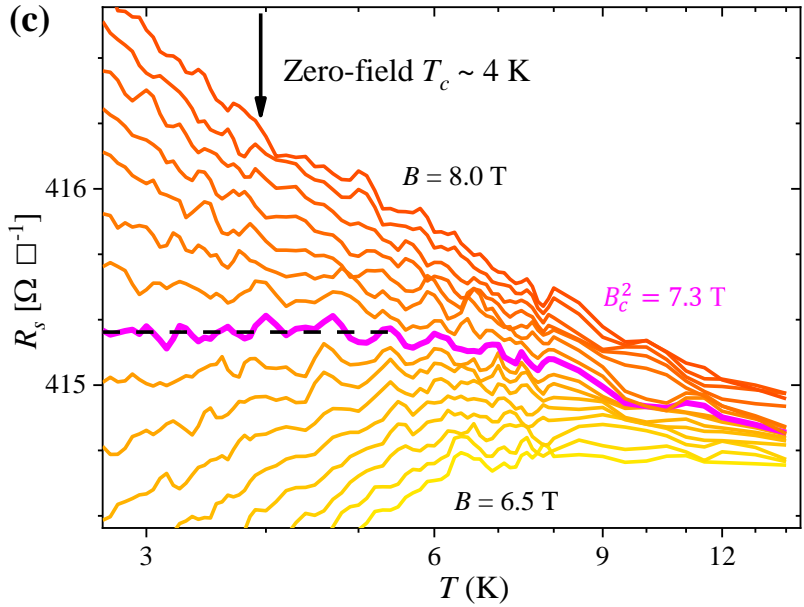

FIG. 1. Magnetic-field driven superconductor to insulator transition. (a) Sheet resistance $R_s$ as a function of temperature for different magnetic fields from 0 to 9 T in steps of 0.5 T. The $R_s(T)$ at the critical field $B_c^1$ is shown as a black line. (b) Detailed $R_s(T)$ data in steps of 0.1 T showing the characteristic magnetic field $B_c^1$, for which $R_s$ is constant from $T \sim 0.6$ K down to the zero-temperature limit. The minima in the $R_s(T)$ data, which define the boundary between Bose insulating state and the normal phase, are marked

with dashed lines. (c) Corresponding $R_s(T)$ data for $B > B_c^1$, showing the characteristic magnetic field $B_c^2$ separating the Bose-insulating state with pronounced superconducting fluctuations, from the Fermi-insulating state. Above $B_c^2$, there is not any signs of superconducting fluctuations.

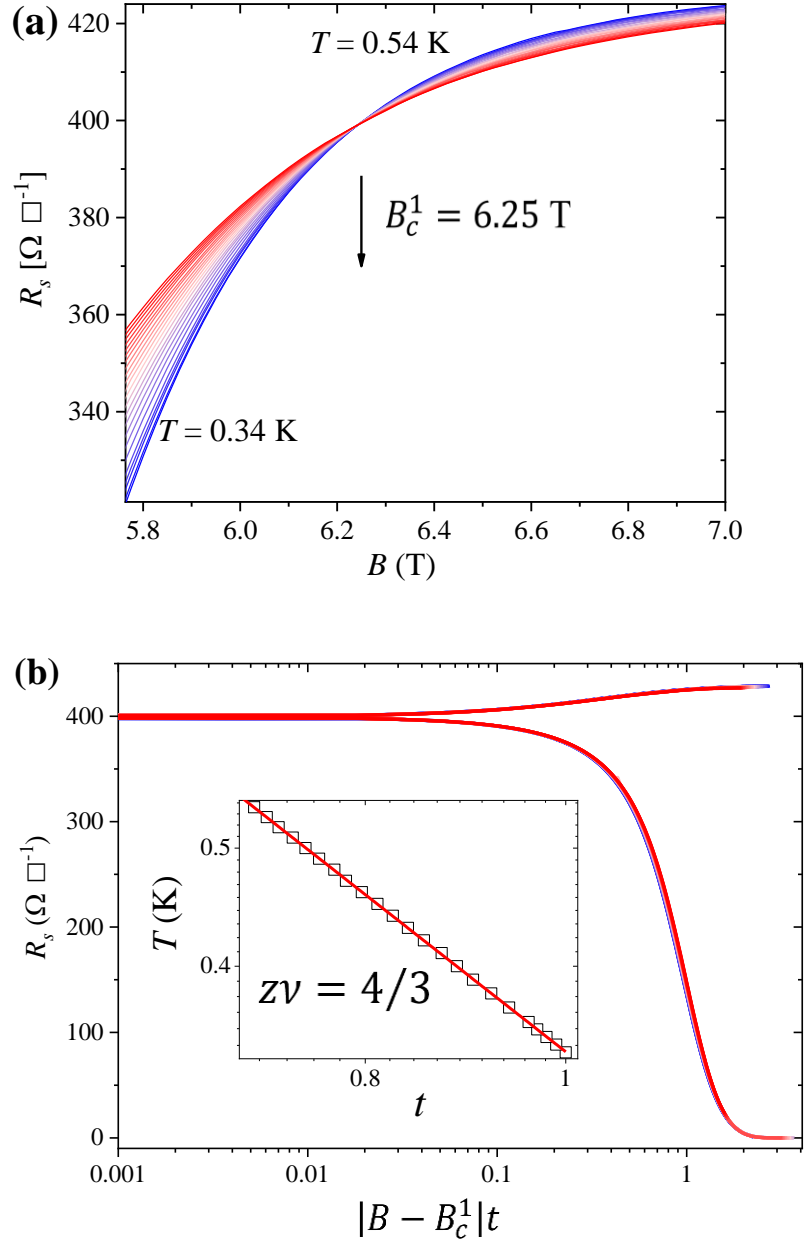

FIG. 2. Scaling analysis at $B_c^1$. (a) Sheet resistance $R_S$ as a function of magnetic field $B$ for different temperatures from 0.34 to 0.54 K. The crossing point is at $B_C^1 = 6.25$ T, $R_C^1 = 399.7$ Ω/□. (b) Scaling-

analysis plot of $R_S$ as a function of $|B - B_c^1|t$. The scaling analysis is performed in fields ranging from 0 to 9 T. Inset: temperature dependence of the scaling parameter $t$, with a power-law fit according to $zv = 4/3$ (solid line).

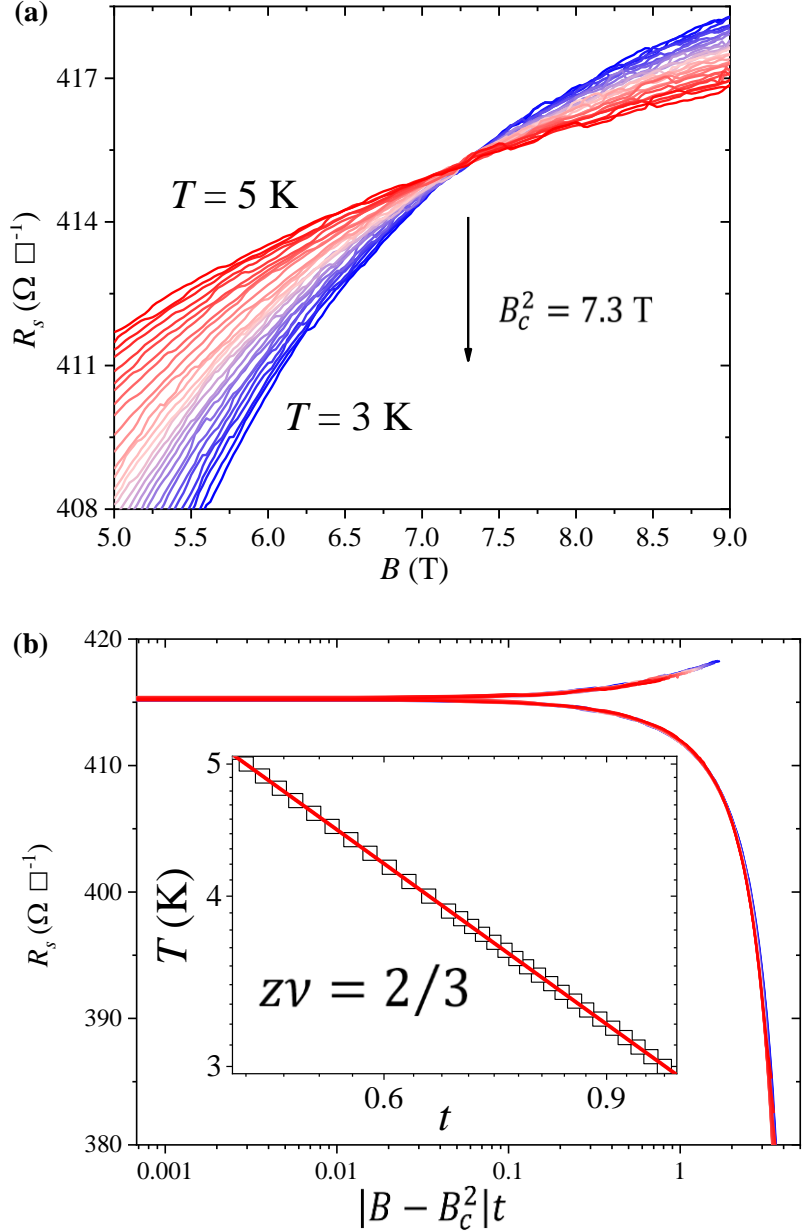

FIG. 3. Scaling analysis at $B_c^2$. (a) Sheet resistance $R_s$ as a function of magnetic field $B$ for different temperatures between 3 to 5 K, with a distinct crossing point at ($B_c^2 = 7.3$ T, $R_c^2 = 415.3$ Ω/□). (b)

Scaling-analysis plot of $R_s$ as a function of $|B - B_c^2|t$ for fields from 4 to 9 T. Inset: temperature dependence of the scaling parameter $t$, with $z\nu = 2/3$.

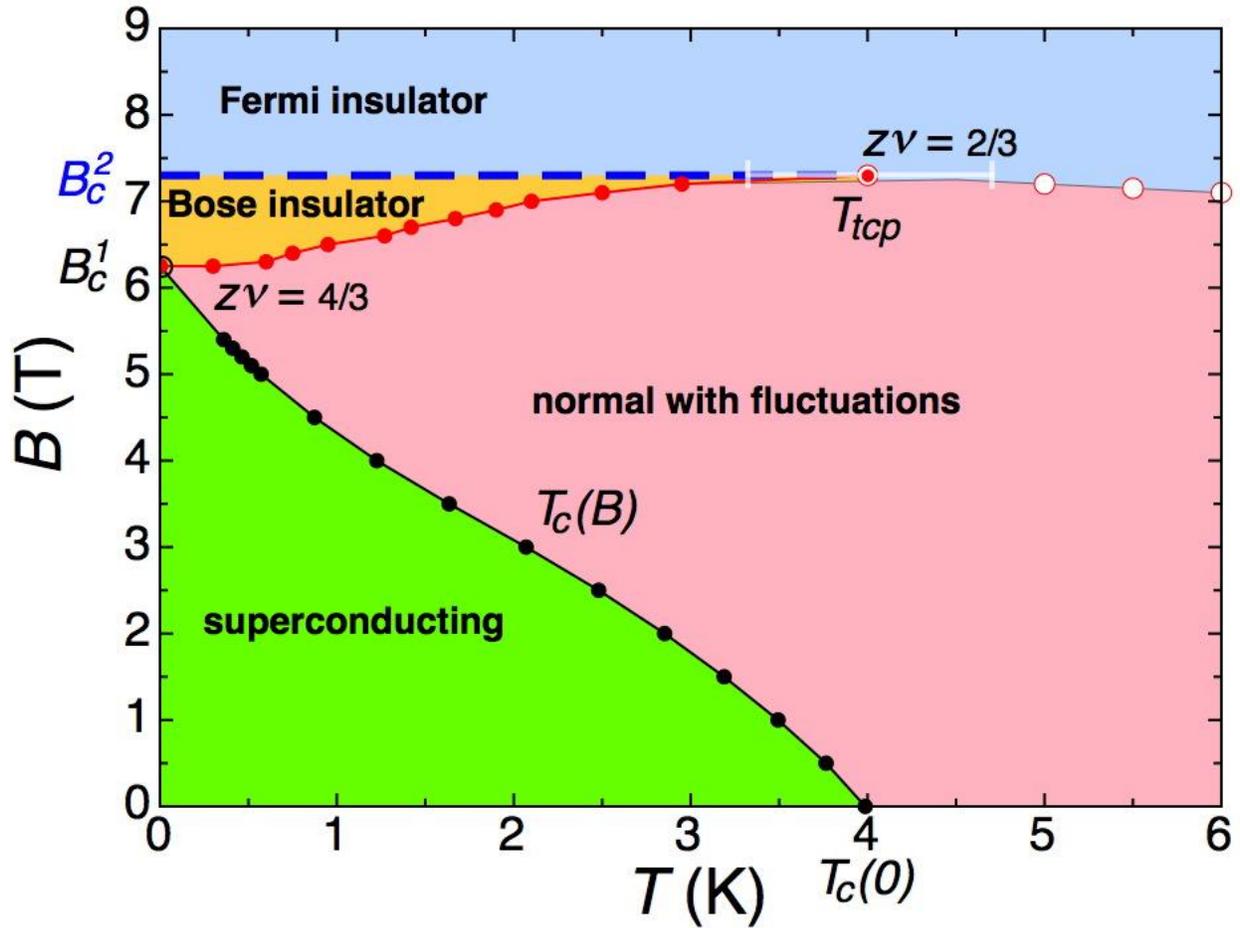

FIG. 4. Sketch of the superconductor to insulator transition in B-T phase diagram. The boundary between the Bose insulating state and the fluctuation-dominated normal state (filled red cirlces) is determined by the minimum of the $R_s(T)$ curves at different fields between $B_c^1$ and $B_c^2$. The separation between the Fermi insulating state and the fluctuation-dominated normal state (open circles) is defined by the maximum of the $R_s(T)$ curves beyond $T_c(0)$. The tricritical point at $T_{\text{tcp}}$ is approximately located at $T_c(0)$, but with a large error margin.